\documentclass[a4paper,10pt]{article}

\title{The Fundamental Constants in Physics}
\author{H. Fritzsch\\University of Munich, Physics Department\\ Munich, Germany}
\begin{document}
\maketitle
\begin{abstract}
We discuss the fundamental constants of Physics in the Standard Model
and possible changes of these constants on the
cosmological time scale. The Grand Unification of the strong,
electromagnetic and weak interactions implies relations between the
time
variation of the finestructure constant $\alpha$ and of the QCD scale
$\Lambda_c$. A change of $\alpha $ by $10^{-15}$ / year, as seen by an
astrophysics experiment, implies thus a time variation of $\Lambda_c$
of
at least $10^{-15}$ / year. An experiment in Quantum Optics at the MPQ
in Munich, which was
designed to look for a time variation of $\Lambda_c$, is discussed.\\
\\
\underline{Contents}
\begin{enumerate}
\item[1)] Introduction
\item[2)] Fundamental Constants and the Standard Model
\item[3)] The Time Variation of the Finestructure Constant
\item[4)] Grand Unification and Time Variation
\item[5)] Results from Quantum Optics
\item[6)] Conclusions and Outlook
\end{enumerate}
\end{abstract}

\section{The Standard Model}
The Standard Model consists of
\begin{enumerate}
\item[a)] the gauge theory of the strong interactions: Quantum
Chromodynamics (QCD)\cite{Fritzsch},
\item[b)] the gauge theory of the electroweak interactions, based on
the gauge group $SU(2) \times U(1)$\cite{Glash}.
\end{enumerate}
QCD is an unbroken gauge theory, based on the gauge group
$SU(3)$, acting in the internal space of ''color``.
The basic fermions of the theory are the six quarks, which form
color triplets. The gluons, the eight massless gauge bosons, are $SU(3)$-- 
octets. The interactions of the quarks and gluons are dictated by the gauge 
properties of the theory. The quarks and gluons interact through the vertex
$g_s \cdot \bar q \gamma_{\mu} \frac{\lambda_i}{2} q \cdot A^{\mu}_i$,
where
$q$ are the quark fields and $A^{\mu}_i$ the eight gluon fields. The
eight
$SU(3)$--matrices are denoted by $\lambda_i$. The strength of the
coupling
constant is
given by  $g_s$.

QCD is a non--Abelian gauge theory. There is a direct
coupling of the gluons among each other. There is also a trilinear coupling,
proportional to $g_s$, and a quadrilinear coupling, proportional to
$g^2_s$.
It is assumed, that the QCD interaction
leads to a confinement of all colored quanta, in particular of the
quarks and
the gluons. But this is thus far not proven.
Replacing the continuous space--time continuum by a lattice, one can
solve the QCD field equations with the computer. The results confirm
the
confinement hypothesis.

The experimental data are in very good agreement with QCD \cite{Rev}.
Quantum Chromodynamics has the property of
asymptotic
freedom. The strength of the quark-gluon-interaction converges to zero
on a logarithmic scale at high energies. At low energies the
interaction
strength is large. Thus the confinement property of QCD might indeed be true.

The equations, describing the renormalization of the coupling constant,
give for $\alpha_s = \frac{g_s \, ^2}{4 \pi}$:
\begin{eqnarray}
\mu \cdot \frac{\partial \alpha_s}{\partial_{\mu}} & = &
- \frac{\beta_0}{2 \pi} \alpha_s^2
- \frac{\beta_1}{4 \pi^2} \alpha_s^3 - \ldots \nonumber \\
\beta_0 & = & 11 - \frac{2}{3} n_f \nonumber \\
\beta_1 & = & 51 - \frac{19}{3} n_f
\end{eqnarray}
($n_f$: number of relevant quark flavors)

Since the interaction is weak at high energies, the quarks and
gluons appear nearly as pointlike objects at small distances. This has
been
observed in the experiments of deep inelastic scattering of electrons,
myons and neutrinos off nuclear targets.

The strong coupling constant at high energies is small, but not zero.
Therefore one expects violations of the scaling behaviour of the
cross-sections.
This has been seen in many experiments. The
value of the QCD coupling constant $\alpha_s = \frac{g_s \, ^2}{4 \pi}$
depends on the energy. One has found in the analysis of scaling
violations\cite{Rev}:
\begin{equation}
\alpha_s \left( M_z^2 \right) \approx 0.1187 \pm 0.002
\end{equation}
($M_z$: mass of the $Z$--boson, $M_z \cong 91.2 \, \,\, {\rm GeV}$).

We can express $\alpha_s (\mu)$ as a function of the scale parameter
of QCD
$\Lambda_c$:
\begin{eqnarray}
\alpha_s (\mu )^{-1} & \approx & \left( \frac{\beta_0}{4 \pi} \right)
ln
\left( \frac{\mu^2}{\Lambda_c^2} \right) \nonumber \\
\beta_0 & = & \left( 11 - \frac{2}{3} n_f \right)
\end{eqnarray}
The experiments give the following value:
\begin{equation}
\Lambda_c \approx 217^{+25}_{-23} \, \, \, {\rm MeV} \, .
\end{equation}

The electroweak gauge theory is based on the gauge group
$SU(2) \times U(1)$. Thus there are three $W$--bosons, related to the
$SU(2)$
group, and a $B$--boson,
related to the $U(1)$--group. The lefthanded quarks and leptons are
$SU(2)$--doublets, the righthanded leptons and quarks are singlets.
Parity is
violated in a maximal way.

The gauge invariance of the $SU(2) \times U(1)$--model is broken
by the ''Higgs``--mechanism\cite{Higgs}. The masses of
the
gauge
bosons are generated by a spontaneous symmetry breaking. Goldstone
bosons
appear as longitudinal components of
the gauge bosons. In the standard ''Higgs`` mechanism there exists a
self--interacting
complex doublet of scalar fields. In the process of symmetry breaking
the
neutral component of the scalar doublet acquires a vacuum expectation
value
$v $, which is determined by the Fermi constant of the weak interactions. 
Therefore the vacuum expectation value is known from the
experiments, if the theory is correct:
\begin{equation}
v \cong 246 \, \, \, {\rm GeV}
\end{equation}

This energy sets the energy scale for the electroweak symmetry breaking. 
Three massless Goldstone bosons are generated, but they  are absorbed to give 
masses to the $W^+, W^-$ and $Z$--bosons. One component of the complex doublet
is not absorbed. This is the ''Higgs``--boson, thus far a hypothetical
particle. It would be the only
elementary scalar boson in the Standard Model. One hopes to find this
particle with the new accelerator LHC at CERN (start in 2009).

In the electroweak model one has two neutral gauge bosons,
which
are mixtures of $W_3$ and $B$, the $Z$--boson and the photon. The
associated
electroweak mixing angle $\Theta_w$ is a fundamental parameter which
has to be fixed by
experiment. It is given by the $Z$--mass, the Fermi constant and the
fine structure constant $\alpha$:

\begin{equation}
sin^2 \Theta_w \cdot cos^2 \Theta_w =
\frac{\pi \alpha \left( M_z \right)}{\sqrt{2} \cdot G_F \cdot M_Z^2} \,
.
\end{equation}

In the experiments one finds $ sin^2 \Theta_w \approx 0.231.$

Note that the electroweak mixing angle is also related to the
mass ratio $M_W / M_Z$. If one neglects radiative corrections, one finds:

\begin{eqnarray}
sin^2 \Theta_w & = & 1 - M_W^2 / M_Z^2 \nonumber \\
M_Z & = & M_W / cos \Theta_w \, .
\end{eqnarray}

In the Standard Model the interactions depend on 28 fundamental
constants.
These are:\\
the constant of gravitiy $G$,\\
the finestructure constant $\alpha $,\\
the coupling constant $g_w$ of the weak interactions,\\
the coupling constant $g_s$ of the strong interactions,\\
the mass of the W-boson,\\
the mass of the ''Higgs``--boson,\\
the masses of the three charged leptons, $m_e, m_{\mu}, m_{\tau}$,\\
the neutrino masses $m(\nu_1), m(\nu_2), m(\nu_3)$,\\
the masses of the six quarks $m_u, m_d, m_c, m_s, m_t, m_b$,\\
the four parameters, describing the flavor mixing of the quarks,\\
and the six parameters, describing the flavor mixing of the leptons,
measured by the neutrino oscillations.

In physics we are dealing with the laws of nature, but little thought
is given to the boundary condition of the universe, related directly to
the Big Bang.
We do not know at the moment, what role is played by the fundamental
costants,
but these constants could form a bridge between the boundary conditions
and the local laws of nature. Thus they would be accidental relics of the 
Big Bang.

Some physicists believe that at least
some of the fundamental constants are just cosmic accidents, fixed by the 
dynamics of the Big Bang. Thus the constants are arbitrary, depending on 
details of the Big Bang. Obviously in this case there is no way to calculate
the  fundamental constants.

Some fundamental constants might be cosmic accidents, but it is
unlikely, that this is the case for all fundamental constants. New
interactions, discovered e. g. with the new LHC--accelerator at CERN,
might offer a way to calculate at least some of the fundamental
constants.

We also do not understand, why the fundamental constants are constant
in
time.
Small time variations are indeed possible and even suggested by
astrophysical experiments. In the theory of superstrings one expects
time
variations of the fundamental constants, in particular of the
finestructure
constant, of the QCD scale parameter $\Lambda_c$, and of the weak
interaction
coupling constant\cite{Tay, Witt}.

If one finds that the fundamental constants are changing in time, then
they
are not just numbers, but dynamical quantities which change according to some 
deeper laws that we have to understand. These laws would be truly fundamental 
and may even point the way to a unified theory including gravity.

\section{Fundamental Constants in the Standard Model}

The Standard Model of particle physics is the theory of the observed particle
physics phenomena. However it depends  on 28 fundamental constants. Within the 
Standard Model there is no way to calculate these constants.

The most famous fundamental constant is the finestructure constant
$\alpha$, introduced in 1916 by Arnold Sommerfeld:
\begin{equation}
\alpha = \frac{e^2}{\hbar c} \, .
\end{equation}
In this constant the electromagnetic coupling $e$ enters, as well as the 
constant of the quantum physics h, and the speed of light $c$.
Sommerfeld realized that $\alpha$ is a dimensionless number, close to
the inverse of the prime number 137. The experiments give the following
value for $\alpha^{-1}$: 137,03599911(46)\cite{Rev}.

Werner Heisenberg proposed in 1936 the relation:
\begin{equation}
\alpha = 2^{-4} \, 3^{-3} \pi \, ,
\end{equation}
which gives $\alpha^{-1} = 137,51$. In 1971 Wyler\cite{Wyler} published
the following expression for $\alpha$:
\begin{equation}
\alpha = \frac{9}{8 \pi ^4} \left( \frac{\pi ^ 5}{2^4 \cdot 5!}
\right)^{1/4}
\, ,
\end{equation}
which gives $\alpha^{-1} = 137,03608$.

Richard P. Feynman wrote about the finestructure constant\cite{Feyn}:
''It has been a mystery ever since it was discussed more than fifty
years ago, and all good theoretical physicists put this number up
on their wall and worry about it. Immediately you would like to know
where this number for a coupling comes from: is it related to $\pi $
or perhaps to the base of the natural logarithms? Nobody knows.
It's one of the greatest mysteries of physics: a magic number that
comes to us with no understanding by man \ldots''.

In quantum field theory the strength of an
interaction is not a fixed constant, but a function of the energy
involved.
The groundstate of a system is filled with virtual pairs of quanta, e.g.
with $e^+e^-$--pairs in QED. Thus an electron is surrountded by
$e^+e^-$--pairs.
The virtual electrons are repelled by the electrons, the virtual
positrons  are attracted. The electron charge is partially shielded by the 
virtual positrons. At relatively large distances the electron charge is smaller 
than at distances less than $\lambda_c$. The dependence on the energy is 
described by the renormalization group equations of Murray Gell--Mann and
Francis Low\cite{Gell}:
\begin{equation}
\frac{d}{d \, ln \left(q/M \right)} e(q) = \beta(e) \, ,
\end{equation}
where
\begin{equation}
\beta(e) = \frac{e^3}{12 \pi ^ 2} + \, \, \, {\rm higher \, \, \, order
\, \, \, terms} \, \, .
\end{equation}

In QED one has to include not only virtual
$e^+e^-$--pairs,
but also the $\mu^+ \mu^-$-- and $\tau^+ \tau^-$--pairs, as well as the
quark--antiquark--pairs. One finds that the finestructure
constant
$\alpha$ at the mass of the $Z$--boson should be the inverse of 128,
in good agreement with the experimental data taken with the
LEP--accelerator\cite{Rev}.

Another fundamental parameter of the Standard Model is the mass of the proton.
In QCD the proton mass is a parameter,
which can be calculated as a function of the QCD scale parameter
$\Lambda_c$ and of the light quark masses. The QCD scale parameter has
been determined many experiments:

\begin{equation}
\Lambda_c = 217 \pm 25 \, \,\, {\rm MeV} \, .
\end{equation}
($\Lambda_c$ is defined in the modified minimal subtraction
$\left( \bar{MS} \right)$ scheme for five quark flavors).

The QCD theory gives a very clear picture of the mass generation. In the
limit, where the quark masses are neglected, the nucleon mass is the
confined field energy of the gluons and quarks. It can be written as:

\begin{equation}
M(Nucleon) = const. \cdot \Lambda_c \, .
\end{equation}
The {\it const.} has been calculated using the lattice approach to QCD.
It is
about 3,9, predicting a nucleon mass in the limit $m_q = 0$ of about
860 MeV.
The observed nucleon mass (about 940 MeV) is higher, due to the
contributions
of the mass terms of the  light quarks $u, d, s$, which in reality are
not
massless.

The mass of the proton can be decomposed as follows:
\begin{eqnarray}
M_p & = & \, \, \, {\rm const.} \, \, \, \Lambda_c + \\
& & < p \mid m_u \bar u u \mid p > + < p \mid m_d \bar dd \mid p >
+ < p \mid m_s \bar s s \mid p > + c_{\rm elm} \, \, \, \cdot \Lambda_c
\, . \nonumber
\end{eqnarray}
The last term describes the electromagnetic self--energy.
It is proportional to the QCD--scale $\Lambda $.
Calculations give\cite{Leut}:
\begin{equation}
c_{\rm elm} \cdot \Lambda_c \approx 2.0 \, \, \, {\rm MeV} \, .
\end{equation}
The up--quark mass term contributes about 20 MeV to the proton mass, the
$d$--quark mass term  about 19 MeV.
Thus the $d$--contribution to the proton mass is about as large as the
$u$--contribution, although there are two $u$--quarks in the proton,
and only
one $d$--quark. This is due to the fact that the $d$--mass is
larger
than the $u$--mass.

In chiral perturbation theory the $u$-- and $d$--masses can be
estimated\cite{Leut1}:
\begin{eqnarray}
m_u & \approx & 3 \pm 1 \, \, \, {\rm MeV} \nonumber \\
m_d & \approx & 6 \pm 1.5 \, \, \, {\rm MeV} \, .
\end{eqnarray}
These masses are normalized at the scale $\mu = 2 \, \, \,
{\rm GeV}$.
Note that quark masses are not the masses of free particles, but of dynamical 
quantities. They depend on the energy scale $\mu $, relevant for the discussion.

The mass of the strange quark can also be estimated in the
chiral perturbation theory\cite{Leut1}.
One finds at $\mu = 2 \, \, \, {\rm GeV}$:
\begin{equation}
m_s \approx 103 \pm 20 \, \, \, {\rm MeV} \, .
\end{equation}
The mass of the strange quark is about 20 times larger than the
$d$--
mass.
Although there are no valence $s$--quarks in the proton, the
$\bar ss$--pairs contribute about 35 MeV to the proton mass, i. e. more
than
the $\bar uu$-- or $\bar dd$--pairs, due to the large ratio $m_s /
m_d$.
Heavy quarks, e. g. $c$--quarks, contribute at most $ \sim $ 1 MeV to
the
nucleon mass\cite{Shif}.

We can decompose the proton mass as follows, leaving out the
contribution
of the heavy quarks:
\begin{eqnarray}
M_p & = & 938 \, \, \, {\rm MeV} \nonumber \\
& = & (862 \qquad  + \qquad 20 \qquad + \qquad 19 \qquad + \qquad 35
\qquad
+ \qquad 2) \, \, \, {\rm MeV} \nonumber \\
& & \, \, \, \, \uparrow \, \, \, \, \, \, \, \, \hspace{2cm} \uparrow
\, \, \, \, \, \hspace{2cm} \uparrow \, \, \, \, \, \hspace{2cm}
\uparrow \, \, \, \hspace{2cm} \uparrow \nonumber \\
& & QCD \, \, \hspace{1cm} u-quarks \, \, \, \hspace{0.5cm}
d-quarks \, \, \, \hspace{0.4cm} s-quarks \, \, \, \, \, \hspace{0.2cm}
QED
\end{eqnarray}
The masses of the heavy quarks $c$ and $b$ can be estimated by
considering
the spectra of the particles, containing $c$-- or $b$--quarks, e. g.
the
charm--mesons or the $B$--mesons. One finds\cite{Rev}:
\begin{eqnarray}
m_c: 1.15 & \ldots & 1.35 \, \, \, {\rm GeV} \, \, \,  (\bar{MS}-mass)
\nonumber\\
m_b: \,  4.1 & \ldots & 4.4 \, \, \, {\rm GeV} \, \, \, (\bar{MS}-mass.
\end{eqnarray}
The dark corner of the Standard Model is the sector of the fermion
masses.
There are the six quark masses, three charged fermion masses, three
neutrino masses, four flavor mixing parameters of the quarks and
six
flavor mixing parameters of the leptons (if neutrinos are Majorana
particles).
These parameters make up 22 of the 28 fundamental constants.

What are the fermion masses? We do not know. They might also be due to
a
confined field energy, but in this case the quarks and leptons would
have
to have a finite radius, as in composite models. The masses would
be
generated by a new interaction. The experiments give a limit on the
internal
radius of the leptons and quarks, which is of the order of $10^{-17}$
cm\cite{Rev}.

In the Standard Model the masses of the leptons and quarks are
generated
spontaneously, like the $W$ and $Z$--masses. Each fermion couples with
a certain strength to the scalar ''Higgs``--boson via a Yukawa
coupling.
A fermion mass is then given by:
\begin{equation}
m(fermion) = g \cdot V \, ,
\end{equation}
where $V$ is the vacuum expectation value of the ''Higgs``--field. For
the
electron this Yukawa coupling constant must be very small, since $V$ is
about 246 GeV:
\begin{equation}
g(electron) = 0,00000208 \, .
\end{equation}
Nobody understands, why this coupling constant is so small. The problem
of
fermion
masses remains to be solved. It seems to be the most fundamental
problem
we are facing at the present time. New experiments at the LHC and at
the
International Linear Collider (ILC) might clarify the issue.

If one is interested only in stable matter, as e. g. in solid state
physics,
only seven fundamental constants enter:
\begin{equation}
G, \Lambda, \alpha, m_e, m_u, m_d, m_s \, .
\end{equation}
The mass of the $s$--quark has been included, since the
$\left( \bar ss \right)$--pairs contribute to the nucleon mass about 40
MeV.
These seven constants describe the atoms and molecules.

It is possible, that there exist relations between the fundamental
constants.
Relations, which seem to work very well, are the relations between the
flavor mixing angles and the quark masses, which where predicted some
time
ago\cite{Frit}:
\begin{eqnarray}
\Theta_u & = & \sqrt{m_u / m_c} \nonumber \\
\Theta_d & = & \sqrt{m_d / m_s} \, .
\end{eqnarray}
Similar relations ca be derived for the neutrino masses and the associated
mixing angles\cite{Frixi}.

These relations are obtained if both for the $u$--type and for the
$d$--type
quarks the following mass matrices are relevant
(texture 0 matrices):
\begin{equation}
M = \left( \begin{array}{lll}
                  0 & A & 0 \\
                  A^* & C & B \\
                  0 & B^* & D
\end{array} \right) \, .
\end{equation}
It would be interesting to know whether such mass matrices are indeed
realized in nature.

\section{Does the Finestructure Constant depend on Time?}

Recent observations in astrophysics\cite{Webb} indicate that the
finestructure
constant $\alpha$ depends on the cosmic time. Billions of years ago it
was
smaller than today. A group of researchers from Australia, the UK and
the USA analysed the spectra of distant quasars, using the Keck
telescope
in Hawaii. They studied about 150 quasars, some of them about 11
billion
lightyears away. The redshifts of these objects varied between 0.5 and
3.5. This corresponds to ages varying between 23\% and 87\% of the age
of
our universe.

They studied the
spectral lines of iron, nickel, magnesium, zinc and aluminium. It was
found
that $\alpha$ is not constant:
\begin{equation}
\frac{\Delta \alpha}{\alpha} = (- 0.72 \pm 0.18 ) \cdot 10^{-5} \, .
\end{equation}
Taking into account the ages of the observed quasars, one concludes
that in
a linear approximation the absolute magnitude of the relative change
of $\alpha $ must be:
\begin{equation}
\left| \frac{d \alpha / dt }{\alpha} \right| \approx 1.2 \cdot 10^{-15}
/ year \, .
\end{equation}
But recent observations of quasar spectra,
performed by different groups, seem to rule out a time variation of
$\alpha $ at the level given above\cite{Chasid, Chaud}.

The idea that the fundamental constants have
a
cosmological time dependence, is not new. In the 1930s P.
Dirac\cite{Dirac}
discussed a time variation of Newtons constant $G$. Dirac argued
that the gravity constant should vary by about a factor of two
during the lifetime of the universe.
The
present limit on the time variation of $G$ is: $\dot{G}/G  \le 10^{-11}
year^{-1}$\cite{Damour}. According to Dirac's hypothesis the time
variation
of $G$ should be about $10^{10} / year$, in conflict with the quoted
limit. In the 1950s L. Landau discussed a possible time variation of the
finestructure constant $\alpha $ in connection with the renormalization
of the electric charge\cite{Land}.

French nuclear physicists discovered that about 1.8
billion
years ago a natural reactor existed in Gabon, West--Africa, close to
the
river Oklo. About 2 billion years ago uranium -235 was more abundant
than
today (about 3,7\%). Today it is only 0,72\%. The water of the river
Oklo
served as a mo\-derator for the reactor. The natural reactor operated
for
about 100 million years.

The isotopes of the rare earths, for example the element Samarium, were
produced by the fission of uranium. The observed distribution of the
isotopes today is consistent with the calculation, assuming that the
isotopes were exposed to a strong neutron flux.

Especially the reaction of Samarium with neutrons is
interesting\cite{Dam}:
\begin{equation}
Sm(149) + n \rightarrow Sm (150) + \gamma \, .
\end{equation}
The very large
cross--section
for this reaction
(about $60 \ldots 90$ kb) is due to a nuclear resonance just above
threshold. The energy of this resonance is very small:
$E = 0.0973 \, \, \, {\rm eV}$. The position of this resonance cannot
have
changed in the past 2 billion years by more than 0.1 eV. Suppose
$\alpha $
has
changed during this time. The energy of the resonance depends in particular 
on the strength of the electromagnetic interaction. Nuclear physics
calculations
give:
\begin{equation}
\frac{ \alpha \left( Oklo) \right) - \alpha (now)}{\alpha (now)} <
10^{-7}
\, .
\end{equation}
The relative change of
$\alpha $ per year must be less than $10^{-16}$ per year, as estimated
by
T. Damour and F. Dyson\cite{Dam}. This conclusion is correct only if no
other
fundamental parameters changed in the past two billion years. If other
parameters, like the strong interaction coupling constant, changed
also,
the constraint mentioned above does not apply.

The Oklo constraint for $\alpha $ is not consistent with the
astrophysical
observation for the relative changes of $\alpha $ of order $10^{-15}$
per year. However, if other parameters also changed in time there will
be
a rather complicated constraint for a combination of these parameters,
but there is no inconsistency.

Recently one has also found a time change of the mass ratio
\begin{equation}
\mu = \frac{M(proton)}{m(electron)} \, .
\end{equation}
One observed the light from a
pair of quasars, which are 12 billion light years away from the
earth\cite{Rein}.
This light was emitted, when the universe was only 1.7 billion years
old.
The study of the spectra revealed, that the mass ratio $\mu $ has
changed in time:
\begin{equation}
\frac{\Delta \mu}{\mu} \approx \left( 2 \pm 0.6 \right) \cdot 10^{-5}
\, .
\end{equation}
Taking into account the lifetime of 12 billion years, the change of
$\mu $
per year would be $10^{-15}$ / year.

\section{Grand Unification}

In the Standard Model we have three basic coupling constants. The gauge
group of the Standard Model is $SU(3)_c \times SU(2) \times U(1)$.

[Show Quoted Text - 660 lines][Hide Quoted Text]
The three gauge interactions are independent of each other.

Since 1974 the idea is discussed that the gauge group of the Standard
Model is a subgroup of a larger simple group. The three gauge
interactions are embedded in a Grand Unified Theory (GUT). A Grand Unification 
implies that $\alpha_3, \alpha_2$ and $\alpha_1$ are related. They can be 
expressed in terms of the unified coupling constant $\alpha_{un}$ and the 
energy scale of the unification $\Lambda_u$.

The simplest theory of Grand Unification is based on the gauge
group
$SU(5)$\cite{Geor}. The quarks and leptons of one generation can be
described by two $SU(5)$--representations. Let us consider the
5--representation of $SU(5)$. After the breakdown of $SU(5)$ to
$SU(3) \times SU(2) \times U(1)$ one obtains:
\begin{eqnarray}
5 & \rightarrow & (3,1) + (1,2) \nonumber \\
\bar 5 & \rightarrow & (\bar 3,1) + (1,2) \, .
\end{eqnarray}
The 5--representation contains a color triplet, which is a singlet
under $SU(2)$, and a color singlet ($SU(2)$--doublet):
\begin{equation}
(\bar 5) = \left( \begin{array}{l}
               \bar d_r\\
               \bar d_g\\
               \bar d_b\\
               \nu_e\\
               e^- \end{array} \right) \, .
\end{equation}
The representation with the next higher dimension is the
10--representation,
which is an antisymmetric second--rank tensor. The 10--representation
decomposes after as follows:
\begin{equation}
(10) \rightarrow (3,2) + (\bar 3,1) + (1,1)
\end{equation}
In terms of the lepton and quark fields of the first generation we can
write the 10--representation (an antisymmetric $5 \times 5$--matrix)
as follows:
\begin{equation}
(10) = \frac{1}{\sqrt{2}} \left( \begin{array}{ccccc}
                 0 & \bar u_b & - \bar u_g & - \bar u_r & - \bar d_r \\
                 - \bar u_b & 0 & \bar u_r & \- \bar u_g & - \bar d_g
\\
                 \bar u_g &  - \bar u_r & 0 & - \bar u_b & - \bar d_b
\\
                 u_r & u_g & u_b & 0 & e^+\\
                 d_r & d_g & d_b & -e^+ & 0 \, .
                 \end{array} \right) \, .
\end{equation}
Combining these two representations, one finds the lepton and
quarks of one generation:
\begin{equation}
\bar 5 + 10 \rightarrow (3,2) + 2 \left( \bar 3,1 \right) + \left(1,2
\right)
+ (1,1) \, .
\end{equation}
For the first generation we have:
\begin{equation}
\bar 5 + 10 \rightarrow \left( {u \atop d} \right)_L + \bar u_L + \bar
d_L
+ \left( {\nu_e \atop e^-} \right)_L + e^+_L \, .
\end{equation}
The second and third generation are analogous. The unification based on
the gauge gruop $SU(5)$ has a number of interesting features:
\begin{enumerate}
\item[1)] The electric charge is quantized.
\begin{equation}
t r Q = O \rightarrow \, Q(d) = \frac{1}{3} \, Q \left( e^- \right)
\end{equation}
\item[2)] At some high mass scale $\Lambda_{un}$ the gauge group of the
Standard Model turns into the group $SU(5)$, and there is only one
single gauge coupling.
The three coupling constants $g_3, g_2, g_1$ for $SU(3), SU(2)$ and
$U(1)$
must be of the same order of magnitude, related to each other by
algebraic constants.
\end{enumerate}
The rather different values of the coupling constants $g_3, g_2, g_1$
at
low energies must be due to renormalization effects. This would also
explain why the strong interactions are strong and the
weak interactions are weak. It is related to the size of the
corresponding
group.

Apart from normalization constants the three coupling constants
$g_3, g_2$ and $g_1$, are equal at the unification mass $\Lambda_{un}$.
Thus the $SU(2) \times U(1)$ mixing angle, given by
$tan \Theta_w = \frac{g_1}{g_2}$, is fixed at or above $\Lambda_{un}$:
\begin{equation}
sin^2 \Theta_w = tr T_3^2 / tr Q^2 = \frac{3}{8} \, .
\end{equation}
At an energy scale $\mu << \Lambda_{un}$ the parameter
$sin^2 \Theta$ changes along with
the three coupling constants:
\begin{eqnarray}
\frac{sin^2 \Theta_w}{\alpha} - \frac{1}{\alpha_s} & = & \frac{11}{6
\pi}
\, ln \, \left( \frac{M}{\mu} \right) \nonumber \\
\alpha / \alpha_s & = & \frac{3}{10} \left( 6 sin^2 \Theta_w - 1
\right) \, .
\end{eqnarray}
At $\mu = M_z$ the electroweak mixing angle has been measured:
$sin^2\Theta_w = 0.2312$.
Note that above the unification energy $\alpha$ and $\alpha_s$ are
related:
\begin{equation}
\alpha / \alpha_s = 3/8 \, .
\end{equation}
This relation can be checked by experiment. In order to get an
agreement between the observed values for $g_3, g_2$ and $g_1$ and the
values predicted by the $SU(5)$ theory, one can easily see that the
unification scale must be very high. Note that
\begin{eqnarray}
ln \left( \frac{M}{\mu}\right) & = & \frac{6 \pi}{11}
\left( \frac{sin^2 \Theta_w}{\alpha} - \frac{1}{\alpha_s} \right)
\nonumber \\
\mu & = & M_Z \nonumber \\
ln \left( M / M_Z \right) & \cong & 39,9 \nonumber \\
M & \approx & 2 \cdot 10^{15} \, \, \, {\rm GeV} \, .
\end{eqnarray}
The precise values of the three coupling constants, determined by
the LEP--experiments\cite{Rev}, disagree with the $SU(5)$ prediction.
The three coupling constants do not converge to a single
coupling constant $\alpha_{un}$\cite{Amal}. A convergence takes place,
if supersymmetric particles are added above the energy of 1 TeV.
Supersymmetry implies that for each
fermion a boson is introduced (s--leptons, s-quarks), and for each boson a 
new fermion is introduced (photino, etc.). These new particles are not
observed in the experiments. It is assumed that they have a mass of about 1 TeV.

The new particles contribute to the renormalization of the gauge
coupling
constants at high energies (about 1 TeV). A convergence of the
three
coupling constants taken place. Therefore a supersymmetric version of 
the $SU(5)$--theory is consistent with the experiments\cite{Amal}.

In theories of Grand Unification like the $SU(5)$--theory one has
quarks, antiquarks and leptons in one fermion representation. Thus
the proton can decay, e. g. $p \rightarrow e^+ \pi^0$. The
lifetime
depends on the mass scale for the unification. For
$\Lambda_{un} = 5 \cdot 10^{14} \, \, \, {\rm GeV}$ in the
$SU(5)$--theory
without supersymmetry one finds $10^{30}$ years for the proton
lifetime.
The experimental lower limit is about $10^{33}$ years.

There is a natural embedding of a group $SU(n)$ into $SO(2n)$,
due to the fact that $n$ complex numbers can be represented by $2n$ real
numbers.
One may consider to use the gauge group $SO(10)$ instead of
$SU(5)$. This was discussed in 1975 by P. Minkowski and the
author\cite{Frimin}. The fermions of one
generation are described by a 16--dimensional spinor representation of
$SO(10)$.

Since $SU(5)$ is a subgroup of $SO(10)$, one has the following
decomposition:
\begin{equation}
16 \rightarrow \bar 5 + 10 + 1 \, .
\end{equation}
The fermions of the $SU(5)$--theory are obtained, plus one
additional
fermion (per family). This state is an $SU(5)$--singlet and describes a
lefthanded antineutrino field. Using the leptons and quarks of the
first
generation we can write the 16--representation as follows in terms of
lefthanded fields:
\begin{equation}
(16) = \left( \begin{array}{lllllllll}
\bar \nu_e & \bar u_r & \bar u_g & \bar u_b & \vdots & u_r & u_g & u_b
&\nu_e \\
e^+ & \bar d_r & d_y & \bar d_b & \vdots & d_r & d_g & d_b &e^-
\end{array} \right)
\end{equation}
A feature of the $SO(10)$--theory is that the gauge group
for the electroweak interactions is larger than in the $SU(5)$--theory.
$SO(10)$ has the subgroup $SO(6) \times SO(4)$. Since $SO(4)$ is
isomorphic into  $SU(2) \times SU(2)$, one finds:
\begin{equation}
SO(10) \rightarrow SU(4) \times SU(2)_L \times SU(2)_R \, .
\end{equation}
The group $SU(4)$ must contain the color group $SU(3)^c$. The
16--representation of the fermions decomposes under $SU(4)$ into
two 4--representations. These contain three quarks
and one lepton, e. g. $\left( d_r, d_g, d_b \right)$ and $e^-$. One may
interpret the leptons as the fourth color. But the gauge group
$SU(4)$  must be broken at high energies (higher than at least 1 TeV):
\begin{equation}
SU(4) \rightarrow SU(3) \times U(1) \, .
\end{equation}
We obtain at low energies the gauge group
\begin{equation}
SU(3)^c \times SU(2)_L \times SU(2)_R \times U(1) \, .
\end{equation}
But the masses of the gauge bosons for the group $SU(2)_R$ must be
much larger than the observed $W$--bosons, related to the group
$SU(2)_L$.

In the $SU(5)$--theory the minimal number of fermions of the Standard
Model
is included. In the $SO(10)$--theory a new righthanded neutrino is
added. This righthanded fermion is
interpreted as a heavy Majorana particle. A mass for the
lefthanded neutrino is generated by the
''see--saw``--mechanism\cite{Moha}.
Thus in the $SO(10)$--theory the neutrinos are massive, while in the
$SU(5)$--theory they must be massless. The $SO(10)$--theory is more
symmetrical than the $SU(5)$--theory.
It is hard to believe that Nature would stop at $SU(5)$, if Nature
has chosen to unify the basic interactions.

In the $SO(10)$--theory there is one additional free parameter, related
to the masses of the righthanded $W$--bosons. Since righthanded charged
currents are not observed, the masses of the associated $W$--bosons must be 
rather high, at least 300 GeV\cite{Yana}. There is a new parameter $M_R$ in 
the $SO(10)$--theory. It can be chosen such that the coupling constant 
converges at very high energies, without using supersymmetry. If one 
chooses $M_R \sim 10^9 \ldots 10^{11}$ GeV, the convergence occurs.

The idea of Grand Unification leads to the reduction of the fundamental
constants by one. The three gauge coupling constants of the Standard
Model
can be expressed in terms of a unified coupling constant $\alpha_u$ at
the
energy $\Lambda_u$, where the unification takes place. The three
coupling constants $\alpha_s, \alpha_2, \alpha_1$ are replaced by
$\alpha_u$ and $\Lambda_u$.

In a Grand Unified Theory the three coupling constants of the Standard
Model are related to each other. If e. g. the finestructure constant
shows a time variation, the other two coupling constants should also
vary
in time. Otherwise the unification would not be universal in time.
Knowing
the time variation of $\alpha $, one should be able to calculate the
time
variation of the other coupling constants.

We shall investigate here only the time change of the QCD coupling constant 
$\alpha_s $.

We use the supersymmetric $SU(5)$--theory to study the time change of
the
coupling constants\cite{Chang, Cal}. The change of $\alpha $ is traced
back
to a change of the unified coupling constant at the energy of
unification
and to a change of the unification energy. These changes are related to
each other:
\begin{equation}
\frac{1}{\alpha} \frac{\dot{\alpha}}{\alpha} = \frac{8}{3} \cdot
\frac{1}{\alpha_s} \cdot \left( \frac{\dot{\alpha}_s}{\alpha_s} \right)
-
\frac{10}{\pi} \frac{\dot{\Lambda}_{un}}{\Lambda_{un}}.
\end{equation}

We consider the following three scenarios:
\begin{enumerate}
\item[1)]
$\Lambda_{un}$ is kept constant, $\alpha_u = \alpha_u(t)$.
We obtain:
\begin{equation}
\frac{1}{\alpha } \frac{\dot \alpha}{\alpha} = \frac{8}{3}
\frac{1}{\alpha_s} \frac{ \dot{\alpha}_s}{\alpha_s} \, .
\end{equation}

Using the experimental value $\alpha_s \left( M_Z \right)
\approx 0.121$, we find for the time variation of the QCD
scale\cite{Chang}:\\
\begin{eqnarray}
\frac{\dot \Lambda}{\Lambda} & \approx & R \cdot
\frac{\dot{\alpha}}{\alpha}
\nonumber \\
\nonumber\\
R & \approx & 38 \pm 6 \, .
\end{eqnarray}
The uncertainty in $R$ comes from the uncertainty in the determination
of the strong interaction coupling constant $\alpha_s$.
A time variation of the QCD scale $\Lambda$ implies a time change of
the
proton mass and of the masses of all atomic nuclei. The change of the
nucleon mass during the last 10 billion years amounts to about 0.3 MeV.

In QCD the magnetic moments of the nucleon and of the atomic nuclei are
inversely proportional to the QCD scale parameters $\Lambda $. We
find
for the nuclear magnetic moments:
\begin{equation}
\frac{\dot{\mu}}{\mu} =
\frac{\frac{d}{dt} \left(\frac{1}{\Lambda} \right)}{\Lambda} =
- \frac{\dot{\Lambda}}{\Lambda} = - R \cdot \frac{\dot{\alpha}}{\alpha}
\, .
\end{equation}
Taking the astrophysics result for $\left( \dot{\alpha} / \alpha
\right)$,
we obtain:
\begin{equation}
\frac{\dot{\Lambda}}{\Lambda} \approx 4 \cdot 10^{-14} / yr \, .
\end{equation}
\item[2)] The unified coupling constant is kept invariant, but
$\Lambda_{un}$
changes in time. In that case we find\cite{Cal}:
\begin{equation}
\frac{\dot{\alpha}}{\alpha} \cong - \alpha \cdot \frac{10}{\pi}
\frac{\dot{\Lambda}_{un}}{\Lambda_{un}}
\end{equation}
and
\begin{equation}
\frac{\dot{\Lambda}}{\Lambda} \approx - 31 \cdot
\frac{\dot{\alpha}}{\alpha}
\, .
\end{equation}
The change of the unification mass scale $\Lambda_{un}$ can be
estimated,
using as input the time variation of the finestructure constant
$\alpha $.
Thus $\Lambda_{un}$ is decreasing
at the rate
\begin{equation}
\dot{\Lambda}_{un} / \Lambda_{un} \approx - 7 \cdot 10^{-14} / yr \, .
\end{equation}
The relative changes of $\Lambda $ and $\alpha $ are opposite in sign.
While
$\alpha $, according to ref. \cite{Webb}, is increasing with a rate of
$10^{-15} / yr$, the QCD scale $\Lambda $ and the nucleon mass are
decreasing with a rate of about $3 \cdot 10^{-14} / yr$. The magnetic
moments
of the nucleons and of nuclei would increase:
\begin{equation}
\frac{\dot{\mu}}{\mu} \approx 3 \cdot 10^{-14} / yr \, .
\end{equation}
\item[3)] The third possibility is that both $\alpha_u$ and
$\Lambda_{un}$
are time--dependent. In this case we find:
\begin{equation}
\frac{\dot{\Lambda}}{\Lambda} \cong 46 \cdot
\frac{\dot{\alpha}}{\alpha}
+ 1,07 \cdot \frac{\dot{\Lambda}_{un}}{\Lambda_{un}} \, .
\end{equation}
On the right two relative time changes appear:
$\left( \dot{\alpha} / \alpha \right)$ and $\left( \dot{\Lambda}_{un} /
\Lambda_{un} \right)$. These two terms might conspire in such a way
that
$\left( \dot{\Lambda} / \Lambda \right)$ is smaller than about
$\left(\pm 40 \cdot \dot{\alpha} / \alpha \right)$.

The question arises, whether a time change of the QCD scale para\-meter
could be
observed in the experiments. The mass of the proton and the masses of
the
atomic nuclei as well as their magnetic moments depend linearly on the
QCD
scale. If this scale changes, the mass ratio $M_{p} / m_e = \mu$ would
change as well, if the electron mass is taken to be constant.

The mass ratio $\mu $ seems to show a time
variation
-- in a linear approximation one has about
\begin{equation}
\frac{\Delta \mu}{\mu} \approx 10^{-15} / year \, .
\end{equation}
If we take the electron mass to be constant in time, this would imply
that
the QCD--scale $\Lambda $ changes with the rate
\begin{equation}
\frac{\Delta \Lambda}{\Lambda} \approx 10^{-15} / year \, .
\end{equation}
\end{enumerate}
The connection between a time variation of the finestructure
constant
and of the QCD scale, discussed above, is only valid, if either the
unified
coupling constant or the unification scale depends on time, not both.
If both the unification scale and the unified coupling constant are
time
dependent, we should use instead eq. (57). There might
be a cancellation between the two terms. In
this case the time variation of the QCD--scale would be smaller than
$10^{-14} / year$. If the two terms cancel exactly, the QCD--scale
would be
constant, but this seems unlikely. Therefore a time variation of the
QCD--scale
of the order of $10^{-15} / year$ is quite possible.

Can such a small time variation of $\Lambda_c$ be observed in the experiments?
In Quantum Optics one can carry out very precise experiments with
lasers. In
the next chapter we shall describe such an experiment at the
Max--Planck--Institute of Quantum Optics in mMunich, which was designed
especially to find a time variation of the QCD scale $\Lambda_c $.

\section{Results from Quantum Optics}

The hydrogen atom is a very
good
test object for checking fundamental theories. Its atomic properties
can be
calculated with very high accuracy. The level
structure of the hydrogen atom can be very accurately probed, using
spectroscopy methods in the visible, infrared and ultraviolett
regions.
Thus the hydrogen atom plays an important r\^ole in determining the
fundamental constants like the finestructure constant.

Measurements of the Lamb shift and the 2S hyperfine structure permit
very
sensitive tests of quantum electrodynamics. Combining optical frequency
measurements in hydrogen with results from other atoms, stringent upper
limits
for a time variation of the finestructure constant\cite{Lang} and of
the
QCD scale parameter can be derived.

The employment of frequency combs\cite{Kola} turned high--precision
frequency
measurements into a routine procedure. The high accuracy of
the
frequency comb have opened up wide perspectives for optical atomic
clock
applications in fundamental physics. Frequency measurements in the
laboratory have become competitive recently in terms of sensitivity to
a
possible time variation of the fine--structure constant.
Though the time interval covered by these measurements is restricted to
a few
years, very high accuracy compensates for this disadvantage. Their
sensitivity becomes comparable with astrophysical and geological
methods
operating on a billion--year time scale.

Important advantages of the laboratory experiments are: The variety of
different systems that may be tested, the possibility to change
parameters
of the experiments in order to control systematic effects, and the
determination of the drift rates from the measured data.
Modern precision frequency measurements deliver information about
the stability of the present values of the fundamental constants, which
can
only be tested with laboratory measurements. But only
non-laboratory methods are sensitive to processes that happened in the
early
universe, which can be much more severe as compared to the present
time.

In the experiment of the MPQ--group in Munich\cite{Lang} one was able
to
determine the frequency of the hydrogen 1S--2S--transition to
2466061102474851(34) Hz.
A comparison with the experiment performed in 1999 gives an upper limit
on
a time variation of the transition frequency in the time between the
two
measurements, 44 months apart. One finds for the difference
$(-29 \pm 57) Hz$, i. e. it is consistent with zero.

The hydrogen spectrometer can be interpreted as a clock, like the cesium clock. 
However in the hydrogen spectrometer one uses a normal transition for the 
determination of the flow of time. This transition depends on the mass of 
the electron and on the fine structure constant. In a cesium clock the flow 
of time is determined by a hyperfine transition, which depends on the fine 
structure constant, but also on the nuclear magnetic moment.

Comparing the $1S - 2S$ hydrogen transition with the hyperfine transition
of Cesium $^{133} Cs$, one can
obtain information about the time variation of the ratio $\alpha /
\alpha_s$.
The Cesium hyperfine transition depends on the
magnetic moment of the Cesium nucleus, and the magnetic moment is
proportional to ($ 1 / \Lambda_c$, ($\Lambda_c$: QCD scale parameter).
If $\Lambda_c$ varies in time, the magnetic moment will also vary.

One has obtained a limit for the time variation of the magnetic
moment of the Cesium nucleus\cite{Lang}:
\begin{equation}
\frac{\delta \mu}{\mu} = \left( 1.5 \pm 2.0s \right) \cdot 10^{-15} /
yr \, .
\end{equation}
These results are consistent with zero. The limit on the time
variation of $\alpha $ is of the same order as the astrophysics result.

The result concerning the magnetic moment implies a limit on the time
variation of $\Lambda_c$:
\begin{equation}
\frac{\Delta \Lambda_c}{\Lambda_c} = \left( - 1.5 \pm 2.0 \right) \cdot
10^{-15} / yr \, .
\end{equation}
This result is in disagreement with our results, based on the
assumption,
that either $\alpha_u$ or $\Lambda_{un}$ change in time. We obtained
about
$10^{-14} / yr$, which is excluded by this experiment.

The result given above is consistent with no time change for
$\Lambda_c$,
but it also agrees with a small time change of the order of $10^{-15}$
per year. If we assume that the electron mass does not change in time,
such a change of $\Lambda_c$ would agree with the astrophysics result
on the time variation of the ratio M(proton) / m(electron)\cite{Rein}.
Theoretically
we would expect such a time variation, if both $\Lambda_{un}$ and
$\alpha_u$ change in time.

\section{Conclusions and Outlook}

We have summarized our present knowledge about the
fundamental
constants and their possible time variation. Today we do not know how
these constants are generated or whether they might depend on time.
There
might be relations between these constants, e. g. between the flavor
mixing
angles and the fermion masses, or relations between the three coupling
constants, implied by the idea of Grand Unification. This would reduce
the number of basic constants from 28 down to a smaller number, but at
least 18 fundamental constants would still exist.

A possible time variation of the fundamental constants must be rather
slow,
at least for those fundamental constants, which are measured very
precisely,
i. e. the finestructure constant, the QCD--scale $\Lambda $, and the
electron mass. The constant of gravity $G$ is known with a precision
of $10^{-11}$. All other fundamental constants, e. g. the masses of the
other leptons or the masses of the heavy quarks, are not known with a
high
precision. The present limits on the time variation of the finestructure 
constant, the QCD scale or the electron mass are of the order of $10^{-15}$ /
year.
These li\-mits should be improved by at least two orders of magnitude
in the
near future.

If the astrophysics experiments indicate a time variation of the order
of $10^{-15}/ year$, it does not mean that experiments in quantum
optics
should also give such a time variation. It might be that until about
10 billion years after the Big Bang the constants did vary slowly, but
after that they remained constant. No theory exists thus far for a
time variation, and there is no reason to believe that a time
variation should be linear, i. e. $10^{-15} / year$ throughout the
history
of our universe. If the fundamental constants do vary, one would expect
that
the variation
is rather large very close to the Big Bang. In the first microseconds
after
the Big Bang constants like $\alpha $ or $\Lambda_c$ might have changed
by a factor 2, and we would not know.

In cosmology one should consider time variations of fundamental
parameters
in more detail.
Perhaps
allowing a suitable time variation of the constants leads to a better
understanding of the cosmic evolution immediately after the Big Bang.
Allowing time variations might lead
to better cosmological theories and to a better understanding of
particle
physics. Particle physics and cosmology together would give a
unified view on the universe.

\end{document}